\documentclass[aps,prl,twocolumn,showpacs,groupedaddress,floatfix]{revtex4}
\usepackage{graphicx}
\usepackage{color}
\begin{document}

%\preprint{}

\title{Photoassociation of a Bose-Einstein Condensate near a Feshbach Resonance}


\author{M. Junker, D. Dries, C. Welford,
J. Hitchcock, Y. P. Chen and R. G. Hulet}
%\thanks{}
\affiliation{Department of Physics and Astronomy and Rice Quantum
Institute, Rice University, Houston, TX 77251, USA}


\date{\today}

\begin{abstract}

We measure the effect of a magnetic Feshbach resonance (FR) on the
rate and light-induced frequency shift of a photoassociation
resonance in ultracold $^7$Li. The photoassociation-induced loss
rate coefficient, $K_p$, depends strongly on magnetic field, varying
by more than a factor of 10$^4$ for fields near the FR. At
sufficiently high laser intensities, $K_p$ for a thermal gas
decreases with increasing intensity, while saturation is observed
for the first time in a Bose-Einstein condensate.  The frequency
shift is also strongly field-dependent and exhibits an anomalous
blue-shift for fields just below the FR.

\end{abstract}

%\keywords{}

\maketitle

%\section{Introduction and Background}

Increasing interest in producing ultracold molecules by associating
ultracold atoms has motivated an improved fundamental understanding
of association processes. Photoassociation (PA) and
magnetically-tuned Feshbach resonances (FR) are two ways in which
gases of ultracold atoms may be connected to the molecular bound
states of their underlying two-body interaction potentials.  A
Feshbach resonance is realized by tuning a molecular bound state
near the scattering threshold \cite{Tiesinga}. The bound state
perturbs the scattering state which can, in turn, strongly affect
the rate of PA \cite{PA_FR}. The rate of association of a
Bose-Einstein condensate (BEC) is of particular interest, as a
universal regime is predicted at extremely high rates of association
\cite{Javanainen, Goral_Holl, Gasenzer, Naidon_Mas} where the rate
is limited by a universal transient response of the atom pairs
\cite{Naidon}, independent of the underlying microscopic association
process.

Several groups have measured absolute PA rate coefficients
\cite{McKenzie, Prodan, Wester, Kraft, Mickelson}. Of these,
however, only two were performed with a quantum gas. In
ref.~\cite{McKenzie}, the rate of PA of a sodium BEC was found to
increase linearly with laser intensity, without any indication of
saturation. We previously investigated PA of $^7$Li at the critical
temperature for condensation $T_c$ (negligibly small condensate
fraction), and observed saturation at a value consistent with the
maximum two-body collision rate imposed by quantum mechanical
unitarity \cite{Prodan}. Other experiments using thermal gases have
made similar observations \cite{Kraft,Unitarity}. The predicted
universal rate limit remains elusive. Previous experiments have also
measured an intensity-dependent spectral red shift of the PA
resonance \cite{Shift,McKenzie,Prodan}, which has been ascribed to
light-induced coupling to the continuum \cite{Fedichev, Bohn,
Jones}. In this paper, measurements of the PA rate coefficient,
$K_p$, and the spectral shift are presented in the vicinity of a
magnetic FR of $^7$Li for both a thermal gas and a BEC. For fields
near the FR, the measured values of $K_p$ vary by more than a factor
of 10$^4$. By exploiting this huge PA rate enhancement, the first
evidence for saturation of the rate of PA-induced loss in a
condensate is obtained. Furthermore, the FR strongly affects both
the magnitude and sign of the spectral shift.

Atomic $^7$Li in the $|F$=1, $m_F$=1$\rangle$ state is prepared in a
single-beam optical dipole trap, as previously described
\cite{Chen}. A uniform magnetic field can be set between 480--900 G
to access both sides of the FR located near 736 G \cite{MJ}.
Evaporative cooling is achieved by reducing the optical trap beam
intensity.  Either Bose condensates with no discernable thermal
fraction, or thermal gases with $T > T_c$, can be produced depending
on the final optical trap depth.

The PA laser propagates collinearly with the optical trap beam, and
is tuned to resonance (peak loss) with a vibrational level $v$ of
the 1$^3\Sigma^+_g$ excited molecular state \cite{Prodan}. There are
no unresolved rotational or hyperfine levels.  For most of the
measurements we tune to the $v=83$ level, located 60 GHz below the
$2^2S_{1/2}\, + \, 2^2P_{1/2}$ dissociation limit \cite{AbrahamJCP}.
The  $v=83$ level is a good compromise between strong free-bound
overlap and relatively weak off-resonance scattering from the atomic
transition. Excited molecules created by the PA laser decay into
pairs of energetic atoms that predominately escape the trap. The
duration of the PA laser pulse $\tau$ is adjusted at each intensity
$I$ and field to maintain the fractional loss of atoms to
approximately 30$\%$.  The rise and fall times of the PA pulse are
less than 0.3 $\mu$s. The number of remaining atoms following the
laser pulse is measured using on-resonant absorption imaging.

The on-resonance loss-rate coefficient, $K_p$, is defined by the
time evolution of the density distribution: $\dot{n}(t,\textbf{r})$
= $-K_p n^2(t,\textbf{r})$, where any possible time-dependence of
$K_p$ during the PA pulse is averaged over.  For the BEC data
(Fig.~4), $\tau$ is much less than either trap period, and
collisional redistribution during the PA pulse can be neglected. In
this case, we solve for the time-dependent density distribution
analytically and extract $K_p$ by spatially integrating the density
and matching the observed and calculated loss after a time $\tau$
\cite{McKenzie}. For much of the thermal data (Figs.~1 and 3),
however, $\tau$ is comparable to the radial trap period and
redistribution is not negligible. Without accounting for the details
of the time evolution, we take $K_p$ to be the average of two
estimates: the first assumes no redistribution, as for the BEC case,
while the second assumes that the distribution remains Boltzmann
throughout. Due to the small fractional loss, we find the difference
between the two estimates to also be small ($<$25$\%$), and account
for it in the stated uncertainties.

Figure 1 shows the field dependence of $K_p$ on both sides of the FR
for a thermal gas. The data show a pronounced dip in $K_p$ at 710 G,
and a maximum, which is more than $10^4$ times larger, located
between 725--750 G. Measurements could not be made closer to the FR
due to high inelastic collisional loss rates, presumably from a
three-body process. A condensate could not be probed on the
high-field side of the FR due to its instability for $a<0$, but at a
given field on the low-field side $K_p$ for a condensate was found
to be approximately half that of a thermal gas, consistent with
quantum statistics.
\begin{figure}
\begin{center}
\includegraphics[width=1\linewidth]{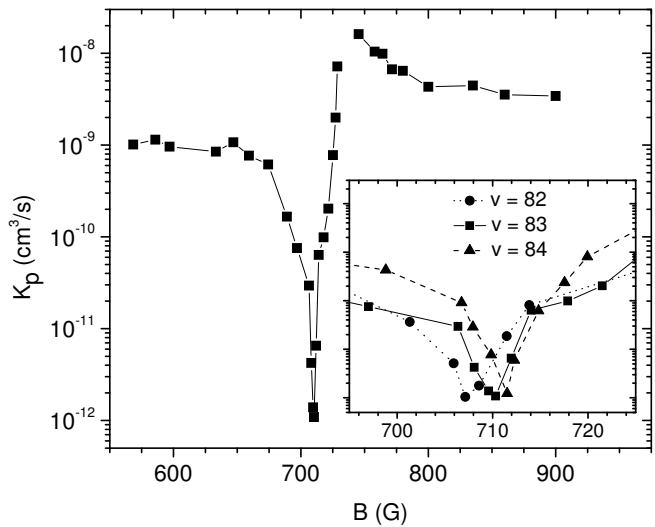}
\caption[Fig. 1]{$K_p$ for PA to $v$=83 for a thermal gas. $I$ is
fixed at 1.65 W/cm$^2$, while $\tau$ is adjusted between 0.07 and
270 ms to keep the fractional loss at $30 \pm 6\%$ ($\tau
\,\propto\, 1/K_p$). The fitted temperatures range between 9--18
$\mu$K. The spread in initial peak density, $0.5-10\times10^{12}$
cm$^{-3}$, reflects a significant decrease in density near the FR,
which we ascribe to enhanced three-body loss. The axial and radial
trapping frequencies are 20 Hz and 3 kHz, respectively. A systematic
uncertainty in $K_p$ of $25\%$ reflects the uncertainty in the
evolution of the density (see text), as well as uncertainties in
probe detuning and intensity. The statistical uncertainty is $20\%$.
The measurement of the highest values of $K_p$ are saturated
(Fig.~3) and are therefore lower limits. The inset shows $K_p$ (same
scale as main figure) for three different excited state vibrational
levels.}
\end{center}
\label{fig:RvsB}
\end{figure}

The variation in $K_p$ can be understood in terms of the
Franck-Condon principle and the dependence of the scattering state
wavefunction $f(R)$ with magnetic field ($R$ is the internuclear
separation). PA predominately occurs at the Condon radius, $R_C$,
defined as the outer classical turning point of the excited
vibrational energy state. Since $K_p \propto |f(R_C)|^2$, a node in
$f(R)$ at $R_C$ results in a minimum in $K_p$. Such minima have been
previously observed \cite{Vuletic}, although with much lower
contrast. For sufficiently low $T$ that the collisional wave vector
$k$ satisfies $k(R-a) \ll 1$, the asymptotic scattering state
$f(R)\,\sim\,k^{1/2}(R-a)$, and a node in $f(R)$ appears at $R
\simeq a$ \cite{Cote}. The inset to Fig.~1 shows that the location
of the $K_p$ minima shift to higher fields (larger $a$) with
increasing $v$, consistent with the fact that $R_C$ increases with
$v$. The peak in $K_p$ near 736 G, on the other hand, is a
manifestation of the enhancement of $f(R_C)$ when $|a|$ is large due
to the FR. The ratio of the asymptotic values of $K_p$ on the high
and low field sides of the FR is $\sim$3.5, which can be understood
from the ratio of $|f(R)|^2$ evaluated using the values of $a$ at
the two asymptotic fields. Using a coupled-channel calculation
\cite{HoubiersPRA98} with previously determined potentials for
lithium \cite{AbrahamPRA97}, we find the scattering lengths at 600 G
and 900 G to be $a_l \simeq 8 \,a_o$ and $a_h \simeq -43 \,a_o$,
respectively, and $R_C \simeq 103 \, a_o$ for $v=83$, where $a_o$ is
the Bohr radius.  With these values, the expected ratio is
$(R_C-a_h)^2/(R_C-a_l)^2 \,\simeq\, 2.4$, in reasonable agreement
with measurement.

The effect of the FR on $f(R)$ is also evident in the light-induced
spectral shifts. We measured spectral shifts for both a thermal gas
and a condensate for fields around the FR. The location of a PA
resonance at a particular field and intensity $I$ is determined by
fitting the resonance lineshape to a Lorentzian. By taking resonance
curves for several values of $I$, we observe that the spectral shift
is linear in $I$, and extract the slope, $\Sigma$, plotted in
Fig.~2. Far from the FR, the shift is red ($\Sigma < 0$), in
agreement with previous measurements of single-photon PA
\cite{Shift, McKenzie, Prodan}. As the FR is approached from low
field, however, $\Sigma$ changes sign, becoming large and positive
just below the FR and large and negative just above it in a
dispersive-like manner. The field where $\Sigma$ vanishes (710 G),
coincides with the location of the zero in $K_p$ shown in Fig.~1.

\begin{figure}
\begin{center}
\includegraphics[bb= 14 16 270 219, clip=true, width=1\linewidth]{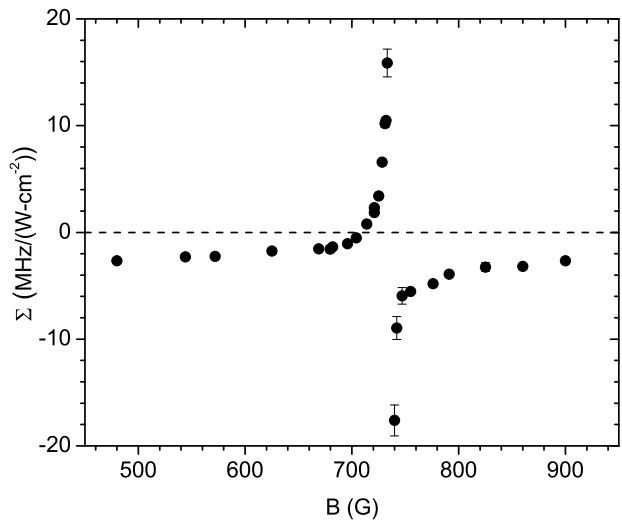}
\caption[Fig. 2]{Slope of the intensity-dependent spectral light
shift $\Sigma$.  The error bars are the standard error of the fitted
slope at each field. For most of the data, this statistical
uncertainty is smaller than the size of the plotted points. The
systematic uncertainty in $\Sigma$ is 10$\%$ due to uncertainty in
$I$. Condensates and thermal gases were found to exhibit the same
$\Sigma$. }
\end{center}
\label{fig:lightshift}
\end{figure}

The spectral shift is a consequence of light-induced mixing of
nearby molecular states into $f(R)$, and has been shown to scale
linearly with $f(R_C)$ \cite{Bohn}. In previous single-photon
experiments the shift was dominated by coupling to the continuum,
resulting in a red shift. In the case of PA near a FR, however, the
shift is strongly affected by the underlying closed-channel
molecular state responsible for the FR. For fields below the FR, the
contribution to the shift from the bound state is positive, while
above resonance it is negative.

The data of Fig.~1 demonstrate that $K_p$ can be extraordinarily
large near the FR, making it an ideal system for exploring
saturation. Figure 3 shows measured values of $K_p$ vs.~$I$ for a
thermal gas at a field just above the FR.  This data can be analyzed
in terms of a Fano model of a bound state coupled to a continuum,
where the on-resonance intensity-dependent loss rate coefficient can
be written as \cite{Bohn}
\begin{equation}\label{scatprob}
K_p(I)= 4K_{max} \frac{I\cdot I_{sat}}{(I+I_{sat})^2},
\end{equation}
where $K_{max}$ is the maximum loss rate coefficient and $I_{sat}$
is the corresponding saturation intensity that accounts for
broadening of the PA resonance. Because of the low temperatures in
our experiment we neglect any energy dependence of $K_p$. The solid
line in Fig.~3 is a fit of the data to Eq.~\ref{scatprob}. The data
clearly show the predicted rollover in $K_p$ that occurs when $I
\,>\, I_{sat}$.  We also observe significant power broadening that
accompanies saturation. By fitting the measured lineshape to a
power-broadened Lorentzian at 35 W/cm$^2$, for example, we obtain
$I_{sat}$ = 4.3 W/cm$^2$, in good agreement with the fit to
Eq.~\ref{scatprob}, which yields 4.1 W/cm$^2$.

\begin{figure}
\begin{center}
\includegraphics[bb= 14 16 270 219, clip=true, width=1\linewidth]{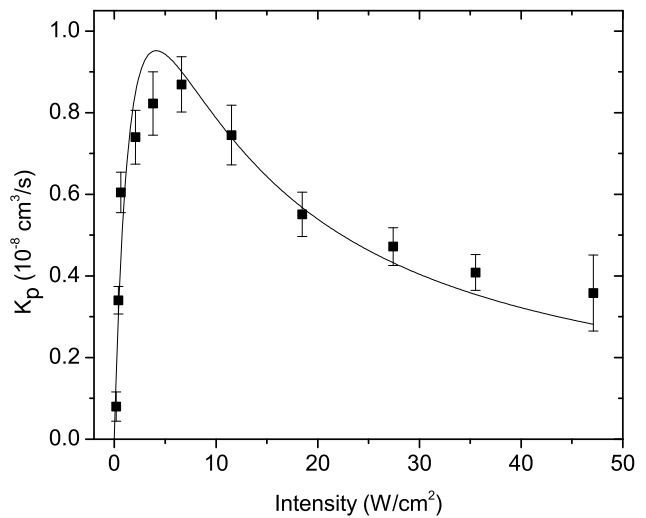}
\caption[Fig. 3]{$K_p$ vs.~$I$ in a thermal gas at 755 G.
$T\,\simeq\, 16$ $\mu$K and the peak density is ~7.5 $\times$
10$^{11}$ cm$^{-3}$. $\tau$ is between 210--250 $\mu$s for all of
the data except at the lowest intensity, for which $\tau$ = 2 ms.
The error bars are the statistical uncertainties from the fitted
$T$. The systematic uncertainties are 30$\%$ for $K_p$ and 10$\%$
for $I$. The solid line is a weighted fit to the data using Eq.~1,
with the fitted parameters $K_{max}$ = 9.5 $\times$ 10$^{-9}$
cm$^3$/s and $I_{sat}$ = 4.1 W/cm$^2$.  A linear fit to the slope
for small $I$ gives 8.2 $\pm$ 1.1 $\times$ 10$^{-9}$ cm$^3$
s$^{-1}$/(W cm$^{-2}$), where the uncertainty refers only to
statistical uncertainty.}
\end{center}
\label{fig:thermrate}
\end{figure}

The theory of ref.~\cite{Bohn} provides an estimate of the
unitarity-limited maximum rate coefficient $K_u \,=\,
2k_BT\Lambda^3/h$, where $h$ is Planck's constant and $\Lambda$ is
the thermal de Broglie wavelength using the reduced mass, and the
factor of 2 accounts for losing both atoms per PA event. For $T$ =
16 $\mu$K, $K_u$ = 8.5 $\times$ 10$^{-9}$ cm$^3$/s, in excellent
agreement with the observed maximum.

Figure 4 shows the measured intensity dependence of $K_p$ for a BEC
near the FR.  By achieving extremely large FR-enhanced loss rates,
saturation was observed in a condensate for the first time. The
maximum $K_p$ of 1.4 $\times$ 10$^{-7}$ cm$^3$/s is nearly a factor
of 10 larger than for any previous PA rate measurement
\cite{Prodan}.

\begin{figure}
\begin{center}
\includegraphics[bb= 14 16 270 219, clip=true, width=1\linewidth]{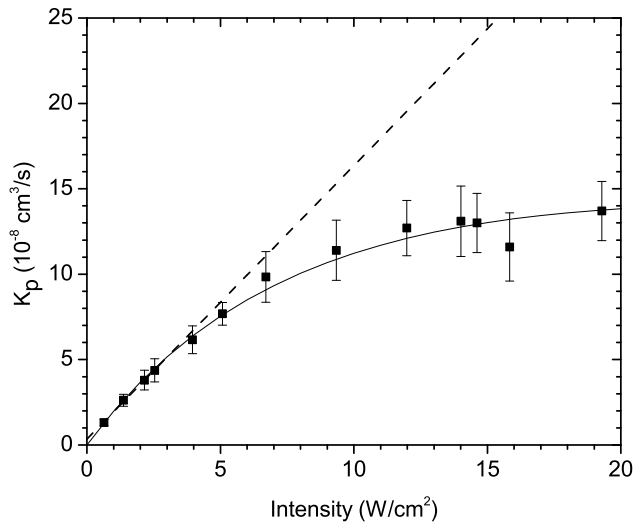}
\caption[Fig. 4]{$K_p$ for a BEC at 732 G ($a\, \simeq\, 1000\,
a_o$). The $1/e^2$ intensity radius of the PA beam of 560 $\mu$m is
much larger than the radial Thomas-Fermi radius of the BEC.  The
pulse duration $\tau \,\propto\, 1/K_p$, is 3 $\mu$s for the highest
$I$ and 50 $\mu$s for the lowest. The peak density is 1.6 $\times$
10$^{12}$ cm$^{-3}$. The axial and radial trapping frequencies are
4.5 Hz and 200 Hz, respectively. The solid line is a fit to Eq.~1
with fit parameters $K_{max}$ = 1.4 $\times$ 10$^{-7}$ cm$^3$/s and
$I_{sat}$ = 26 W/cm$^2$. A linear fit to the low intensity data
gives a slope of 1.6 $\times$ $10^{-8}$ cm$^3$ s$^{-1}$/(W
cm$^{-2}$), as indicated by the dashed line. The error bars are the
statistical uncertainties, due mainly to the measured Thomas-Fermi
radius of the initial density distribution. A systematic uncertainty
of 12$\%$ reflects the uncertainty in the image probe detuning and
intensity. $I$ could not be increased further without producing loss
from dipole forces.}
\end{center}
\label{fig:BECrate}
\end{figure}

A complete theoretical understanding of PA of a BEC in the
high-intensity regime is still evolving, but a framework for
comparison of the data may be established by defining $K_p$ in terms
of a characteristic length, $L$, as $K_p$ = $(\hbar /m)L$. The
smallest relevant length scale is the average interatomic
separation, evaluated at the peak density $n_o$. Taking $L$ =
$n_o^{-1/3}$, we obtain the ``rogue photodissociation" limit,
$K_{pd}$, a universal regime where saturation in the rate of
condensate loss is predicted due to the dissociation of excited
molecules into the hot pair continuum \cite{Javanainen}. The NIST
BEC experiment achieved $K_p \,\simeq\, K_{pd}$, but with no
indication of saturation \cite{McKenzie}.  For the data of Fig.~4,
$n_o$ = 1.6 $\times$ 10$^{12}$ cm$^{-3}$, giving $K_{pd}\,\sim$ 8
$\times$ 10$^{-9}$ cm$^3$/s. Our measured $K_{max}$ is nearly 20
times greater than $K_{pd}$. More recent calculations
\cite{Gasenzer, Naidon}, however, have shown that while dissociation
does impose a rate limit on condensate loss, it is not as stringent
as $K_{pd}$. While the highest-$I$ data of Fig.~4 are well into the
``dissociation regime" \cite{Naidon_Mas, Naidon}, our measured
$K_{max}$ is, nonetheless, nearly 7 times greater then predicted
from the equations given in ref.~\cite{Naidon}. We note that in the
experiment atoms removed from the condensate but left with energies
below the trap depth (500 nK) may be indistinguishable from
condensate atoms by our detection method. This effect cannot explain
the discrepancy with theory, however, as it would result in an
apparent suppression of $K_p$ at the onset of photodissociation.

The observed saturation could be explained by a higher than expected
limit imposed by dissociation, perhaps due to cross-coupling between
the PA and Feshbach resonances \cite{Mackie}. An alternative
explanation is provided by quantum mechanical unitarity. If we take
$L \sim 2R_{TF}$, where $R_{TF}\simeq$~10~$\mu$m is the radial
Thomas-Fermi radius, then $K_{u}\sim$ 1.8 $\times$ 10$^{-7}$
cm$^3$/s, in good agreement with the measured value of 1.4 $\times$
10$^{-7}$ cm$^3$/s. A preprint has recently appeared that directly
models our experiment with good quantitative results \cite{Mackie},
but it does not determine whether the rate limit is imposed by
dissociation or unitarity.

We have presented well-characterized measurements of PA near a FR,
which we hope will serve to constrain and guide theory, especially
in the previously unexplored strong coupling regime.  Condensate
loss rates as high as the unitarity limit have been demonstrated by
combining PA and Feshbach resonances. Future studies should explore
the dynamical behavior of loss since $K_p$ is predicted to be
time-dependent in the dissociation regime \cite{Javanainen,
Gasenzer, Naidon_Mas, Naidon}. Finally, Feshbach enhanced PA may
prove useful in realizing the long-sought goal of coherent
oscillation between atomic and molecular condensates
\cite{HeinzenPRL2000} or for efficient production of ground-state
molecules \cite{Kokkelmans} using a two-photon process
\cite{Two-Photon}.

We thank T. Corcovilos, R.~C\^{o}t\'{e}, T.~Gasenzer, P.~Julienne,
T.~Killian, M.~Mackie, P.~Naidon, G.~Partridge, and H.~Stoof for
valuable discussions. Support for this work was provided by the NSF,
ONR, the Keck Foundation, and the Welch Foundation (C-1133).

% If in two-column mode, this environment will change to single-column
% format so that long equations can be displayed. Use
%\begin{widetext}
%\end{widetext}

% figures should be put into the text as floats.
% Use the graphics or graphicx packages (distributed with LaTeX2e)
% and the \includegraphics macro defined in those packages.
% See the LaTeX Graphics Companion by Michel Goosens, Sebastian Rahtz,
% entire page. There is no need to do explicit centering.

% \begin{figure}
% \end{figure}

% Surround figure environment with turnpage environment for landscape
% figure
% \begin{figure}
% \end{figure}

% Surround table environment with turnpage environment for landscape
% table
% \begin{table}
% \begin{ruledtabular}
% \end{tabular}
% \end{ruledtabular}
% \end{table}

%\appendix
%\section{}


\end{document}